\documentclass[aps,prf,amsmath,amssymb,reprint,superscriptaddress,twocolumn,showpacs,longbibliography]{revtex4-1}

 \usepackage{graphicx}
 \usepackage{amsmath}
 \usepackage{amsfonts}
 \usepackage{amssymb}
 \usepackage{pstricks}
 \usepackage{psfrag}
 \usepackage{epsfig}
 \usepackage{changes}
 \usepackage[ansinew]{inputenc}
\usepackage{mathptmx}
\usepackage{helvet}
\usepackage{textcomp}
\usepackage{ulem}

\begin{document}


\title{Gravity induced formation of spinners and polar order of spherical microswimmers  on a surface}



\author{Zaiyi Shen}
\affiliation{Univ.  Bordeaux, CNRS, LOMA, UMR 5798, F-33405 Talence, France}
\author{Juho S. Lintuvuori}
\email[]{juho.lintuvuori@u-bordeaux.fr}
\affiliation{Univ.  Bordeaux, CNRS, LOMA, UMR 5798, F-33405 Talence, France}


\date{\today}

\begin{abstract}
We study numerically the hydrodynamics of a self-propelled particle system, consisting of spherical squirmers sedimented on a flat surface. We observe the emergence of dynamic structures, due to the interplay of particle-particle and particle-wall hydrodynamic interactions. At low coverages, our results demonstrate the formation of small chiral spinners:  two or three particles are bound together via near-field hydrodynamic interactions and form a rotating dimer or trimer respectively. The stability of the self-organised spinners can be tuned by the strength of the sedimentation. Increasing the particle concentration leads more interactions between particles and the spinners become unstable. At higher area fractions we find that pusher particles can align their swimming directions leading to a stable polar order and enhanced motility.
Further, we test the stability of the polar order in the presence of a solid boundary. We observe the emergence of a particle vortex in a cylindrical confinement.
\end{abstract}


\maketitle


\section{Introduction}
In active materials~\cite{mike2012review,marchetti2013hydrodynamics} the motile constituents can exhibit collective and coherent motion~\cite{vicsek2012collective}. Such collective dynamics are commonly discovered in nature, ranging from various length-scales and species, such as bacterial swarms~\cite{kearns2010field}, fish schools~\cite{filella2018model}, bird flocks~\cite{ballerini2008interaction} and human crowds~\cite{bain2019dynamic}. Understanding the key physical principles behind the collective dynamics can be in addition to the understanding of the dynamic order in nature, also inspire the creation of functional active materials~\cite{yu2018ultra,jin2019mimicking} or robotic applications where the self-cooperation of small machines leads to large coherent assembly capable of carrying out complex tasks~\cite{li2019particle,rubenstein2014programmable}.



On the micrometer length scale, a natural example of active materials is given by bacteria~\cite{kearns2010field,drescher2010direct}, while recently artificial active materials, based on colloidal particles, have become an important tool to study collective motion in laboratory~\cite{zottl2016emergent,bechinger2016active,moran2017phoretic}. A large variety of different collectively moving states have been reported~\cite{wioland2016directed,bricard2015emergent,bricard2013emergence,sumino2012large,lushi2014fluid,grossman2008emergence,thutupalli2011swarming} and geometrical constraint can strongly affect the dynamics~\cite{morin2017distortion,theillard2017geometric,lushi2014fluid,grossman2008emergence,bricard2015emergent,blois2019flow,singh2016universal,schaar2015detention}. Examples of this, from bacterial world, include the hydrodynamic stabilisation of rotating Volvox pairs~\cite{drescher2009dancing} due to a interplay between sedimentation and hydrodynamic effects and the formation of vortex arrays~\cite{petroff2015fast} near confining surfaces, while guiding~\cite{das2015boundaries,simmchen2016topographical} and flow-induced phase separation~\cite{thutupalli2018flow} have been observed with artificial swimmers in confinement.
Potential interactions such as phoretic~\cite{das2015boundaries,simmchen2016topographical,palacci2013living,ginot2018aggregation}, electrical~\cite{yan2016reconfiguring,zhang2016directed}, magnetic~\cite{kaiser2017flocking} in addition to hydrodynamic effects~\cite{shen2019hydrodynamic,petroff2015fast,thutupalli2018flow} can be used to create complex patterns and dynamic self-assembled structures.

In a typical experimental realisation of self-propelling colloids~\cite{palacci2010sedimentation,theurkauff2012dynamic,palacci2013living,ginot2018aggregation}, the particles sediment at the bottom of the container and form a monolayer near the confining surface.

In this work, we want to address the role of hydrodynamic interactions in the emergence of collective dynamics of self-propelled particles sedimenting towards a flat wall. We model the swimmers as a spherical squirmers~\cite{lighthill52} using lattice Boltzmann (LBM) method. 
In recent years the squirmer model has become an important theoretical tool to study the hydrodynamics of self-propelled particles.
For example, a stable polar order has been predicted for weak pullers ($0<\beta<1$) in the bulk~\cite{evans2011orientational,ishikawa2008development,alarcon2013spontaneous} and the presence of a confining surface wall can influence the dynamics of squirmers~\cite{berke2008hydrodynamic,oyama2016purely,schaar2015detention,zottl2016emergent}.
Neutrally buoyant squirmers can be hydrodynamically trapped by a flat no-slip surface~\cite{ignacio1,ignacio2,shen2018hydrodynamic,li2014hydrodynamic,lintuvuori2016hydrodynamic,ishimoto2013squirmer} and near-field hydrodynamic interactions have been shown to play a crucial role~\cite{lintuvuori2016hydrodynamic}.
When the squirmers are subjected to constant gravity, the sedimented particles orient along the wall normal, resulting to a floating state, where the stationary swimmer hovers above the surface and  points directly away from the confining wall~\cite{ruhle2018gravity}.  Constant aligning torque turning the particles towards the wall, has been shown to lead to the dynamic self-assembly of various structures such as the stabilisation of chiral spinners~\cite{shen2019hydrodynamic}, while sedimented particles have been shown to form for example a Wigner fluid~\cite{kuhr2019collective}.

By employing large scale numerical simulations, including near-field hydrodynamic interactions~\cite{lintuvuori2016hydrodynamic} we observe the dynamical structures formed by the squirmers sedimenting at flat surface. The results demonstrate the stabilisation of chiral spinners consisting of hydrodynamically bound dimers and trimers 
and stable polar order for weak pushers. 

\section{Methods}

We study the dynamics of self-propelled particles using lattice Boltzmann simulations. The motile particles are modelled as spherical squirmers~\cite{lighthill52}. The squirmer model does not explicitly deal with phoretic interactions, but instead considers a continuous slip velocity over the particle surface to take into account the different motilities on the surface of the Janus particle~\cite{shen2018hydrodynamic}. The tangential slip velocity at the particle surface {\color{black} is prescribed along the polar direction $\mathbf{e}_{\theta}$ and} is given by~\cite{Magar03}
\begin{equation}
u(\theta)=B_1\sin \theta + B_2 \sin \theta  \cos \theta .
\label{squirmer}
\end{equation}
where $\theta$ is the polar angle with respect to the particle's axis. The bulk swimming speed is given by $u_0= 2 B_1/3$ and the squirmer parameter is $\beta={B_2}/{B_1}$. The $\beta$ defines the hydrodynamic nature of the swimmer: $\beta<0$ corresponds to a pusher and  $\beta>0$ to a puller.

To model the effect of gravity, an external force, in the form of $F_s=6 \pi \mu R u_s$, is applied to make the particles sediment towards the surface. The ratio $u_s/u_0$ between the sedimentation speed $u_s$ and the bulk swimming speed $u_0$ characterises the strength of the sedimentation.

A lattice Boltzmann method~\cite{succi2001lattice} was used to simulate the squirming motion. In the simulations the lattice spacing $\Delta x$, time step $\Delta t$ and the density $\rho$ are set to unity. The boundary condition at the particle surfaces are modified to take into account the surface slip flow (eq.~\ref{squirmer})~\cite{ignacio1,ignacio2}. The fluid viscosity was set to  $\mu = 1/6$ in simulation units and the particle Reynolds number $\rho u_0 R/\mu=0.48$ which is small enough that inertial effects do not play a role in the observed dynamics. The simulations were carried out in a rectangular simulation box with the size of $L_X\times L_Y\times L_Z$, with a no-slip wall at $X=0.5$ and $X=L_X+0.5$ and periodic boundary conditions along $Y$ and $Z$. Unless otherwise mentioned, a particle radius $R = 8$ and $L_X = 6R,~L_Y = 40R$ and $L_Z = 40R$ were used. The simulations were concentrated to reasonably weak pusher/pullers $\beta\in[-2,1]$.

To stop the particles from penetrating each other and the wall, a short range repulsive potential is employed~\cite{lintuvuori2016hydrodynamic,shen2019hydrodynamic}
\begin{equation}
  V(d) = \epsilon\left(\frac{\sigma}{d}\right)^\nu
  \label{soft}
\end{equation}
which is cut-and-shifted by
{\color{black}
\begin{equation}	
	V_W(d) =
	\begin{cases}
	V(d) - V(d_c) - (d-d_c) \frac{\partial V(d)}{\partial d}\mid_{d=d_c}, & d<d_c\\
	0, & d\geq d_c
	\end{cases}
  \label{repulsion}
\end{equation}
}
to ensure that the potential and the resulting force go smoothly to zero at the interaction range $d_c$. The $d$ is defined as the distance between the particle bottom and the surface (as $h$ in the inset in Fig. \ref{spinner}a) or the distance between two particle surfaces. The $\epsilon = 0.6$ and $\sigma = 1.0$ are constant in the reduced units of energy and length, respectively. The $\nu = 12$ controls the steepness of the repulsion.

\section{Results}
\subsection{Hydrodynamic stabilisation of chiral spinners at low surface coverage}
\begin{figure}[tbhp!]
\centering
\includegraphics[width=0.5\textwidth]{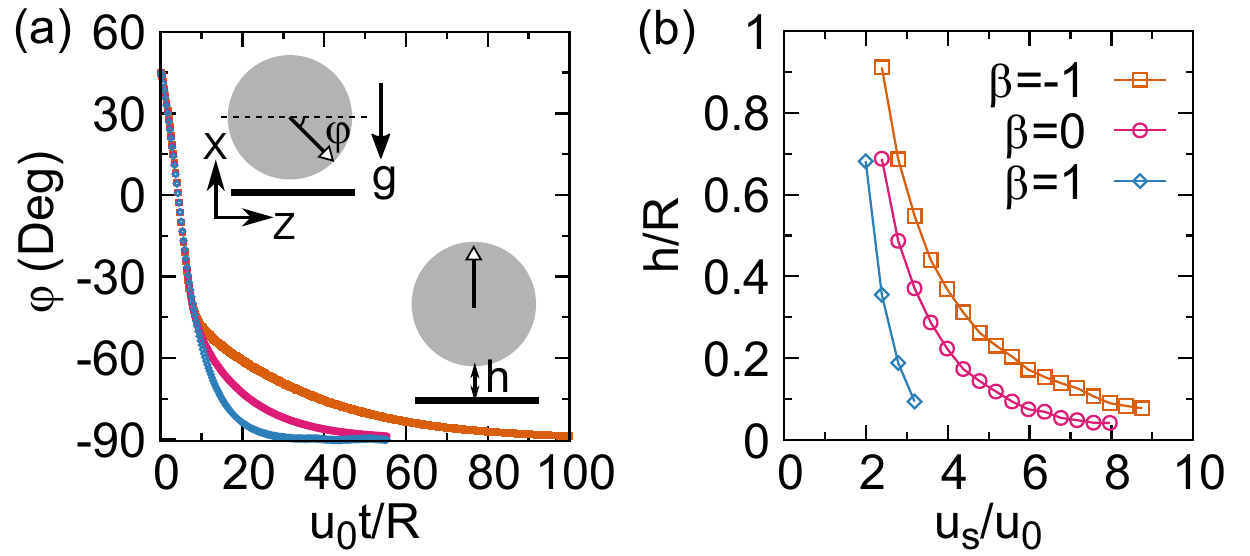}
\caption{\footnotesize {\color{black}(a) The time evolution of the angle $\varphi$ between the particle axis and the wall plane when the particle sediments towards the surface. The sedimentation strength is $u_s/u_0=2.4$. The inset shows the definition of $\varphi$ and the gap-size $h$ used in the text. (b) Floating of a self-propelled particle near the surface. The final stable distance $h$ between the particle surface and flat wall is plotted as a function of the sedimentation strength. The simulation box is $L_X = 12R,~L_Y = 12R$ and $L_Z = 12R$.}}
\label{spinner}
\end{figure}

In order to study in detail the hydrodynamic interactions of sedimenting squirmers near the flat wall,  large particles $R=12\Delta x$ were used. This allows the realisation of the particles at very high resolution~\cite{kuron2019lattice} on the lattice, as well as ensure that external particle-wall repulsion plays no role.
Similarly to ref.~\cite{ruhle2018gravity} we observe a floating state where {\color{black} isolated particles} orient along the wall normal, pointing upright away from the surface, opposite to the direction of the sedimentation force. Fig.~\ref{spinner}a shows the time evolution of the angle $\varphi$ between the wall plane and the swimmer direction (inset in Fig.~\ref{spinner}a), for a weak pusher ($\beta=-1$), a neutral swimmer ($\beta=0$) and a weak puller ($\beta= +1$) when the sedimentation is reasonably weak $u_s/u_0=2.4$.
Initially the particles are aligned towards  the wall with $\varphi = 45^\circ$. The particle slides along the surface and reorients such that in the steady state the particle axis points to the opposite direction to the gravity $\varphi\approx -90^\circ$ (Fig.~\ref{spinner}a). In the steady state, the competition between the sedimentation force and the hydrodynamic self-propulsion, leads to the floating of the particle above the wall with a well defined gap-size $h$ between the particle surface and the confining wall (inset in Fig.~\ref{spinner}a). 

\begin{figure}[tbhp!]
\centering
\includegraphics[width=0.5\textwidth]{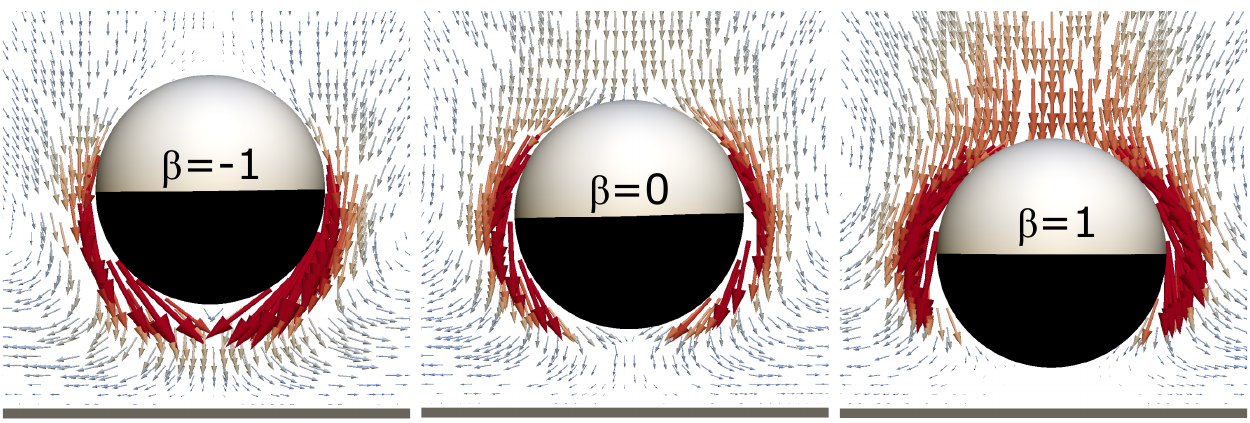}
\caption{\footnotesize {\color{black} The calculated flow fields for a $\beta = -1$ pusher, a neutral swimmer ($\beta=0$) and for a $\beta = +1$ puller floating above a no-slip surface ($u_s/u_0\approx 2.4$). The colour code corresponds to the magnitude of the fluid flow.}}
\label{Single}
\end{figure}

The steady state gap-size $h$ (inset in Fig.~\ref{spinner}a) in the floating state is controlled by both the sedimentation strength and the squirmer parameter $\beta$ (Fig.~\ref{spinner}b). When the sedimentation is reasonably weak
$u_s/u_0\approx 2$, the squirmers float close to their radius ($h\approx R$) away from the wall (Fig.~\ref{spinner}b). 
Generally, these results show good agreement with the simulations using multi particle collision dynamics (MPCD) method~\cite{ruhle2018gravity}. {\color{black} The tendency of the  pushers to float further away from the wall comparing to pullers can be captured by far-field hydrodynamic effects~\cite{ruhle2018gravity} and qualitatively understood by considering the swimming mechanism.  Pushers create a stronger flows behind the particle while with pullers the flow is more pronounced on the top hemisphere. The near-field effects rotate the weak ($|\beta| \leq 1$) squirmers to point directly away from the wall in a steady state~\cite{ruhle2018gravity}. In this state, the pushers create the largest flows near the wall at the back of the particle, while pullers mix the fluid more strongly at the top of the particle farther away from the confining surface (Fig~\ref{Single}).}

\begin{figure}[tbhp!]
\centering
\includegraphics[width=0.5\textwidth]{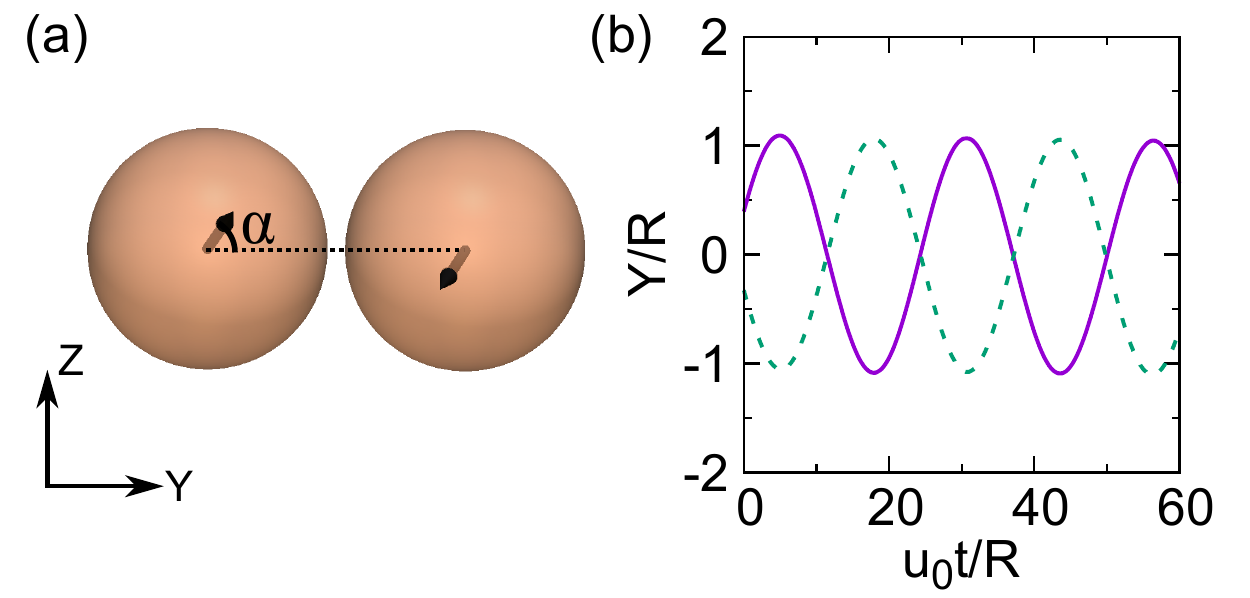}
\caption{\footnotesize {\color{black}(a) A typical configuration of rotating dimer with $u_s/u_0=6.4$ for $\beta=-1$ pushers. The definition of the chirality $\alpha$ is shown in the figure. The black arrows show the projections of the particle axis in $Y-Z$ plane. (b) The $Y$ positions of each particle in the dimer are plotted as a function of time $u_0 t/R$.}}
\label{dimer_exp}
\end{figure}

\begin{figure}[tbhp!]
\centering
\includegraphics[width=0.5\textwidth]{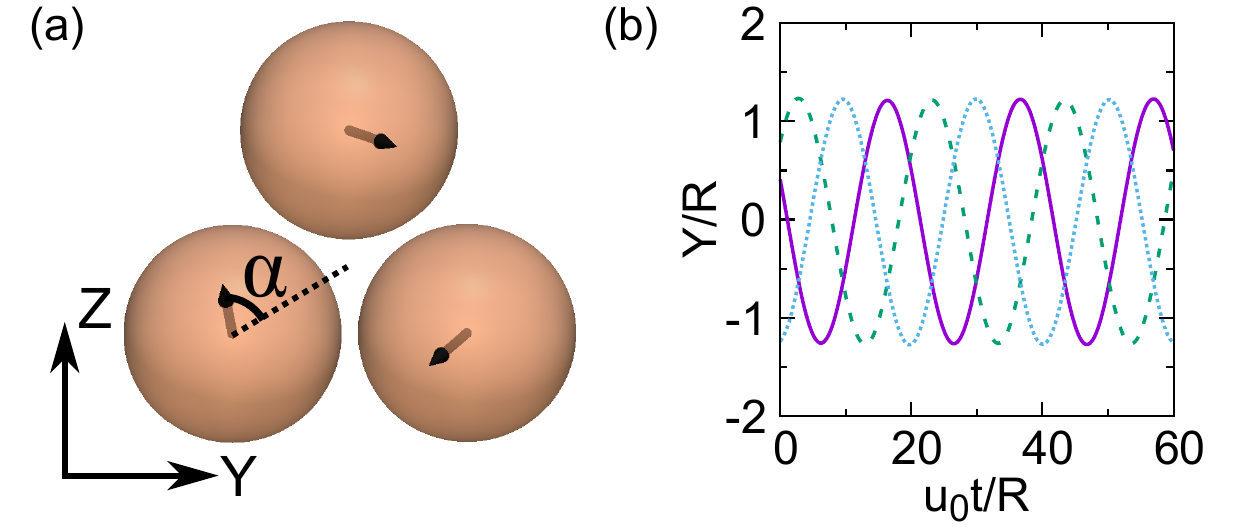}
\caption{\footnotesize {\color{black}(a) A typical configuration of rotating trimer with $u_s/u_0=8$ for $\beta=-1$ pushers. The definition of the chirality $\alpha$ is shown in the figure. The black arrows show the projections of the particle axis in $Y-Z$ plane. (b) The $Y$ positions of each particle in the trimer are plotted as a function of time $u_0 t/R$.}}
\label{trimer_exp}
\end{figure}

\begin{figure}[tbhp!]
\centering
\includegraphics[width=0.5\textwidth]{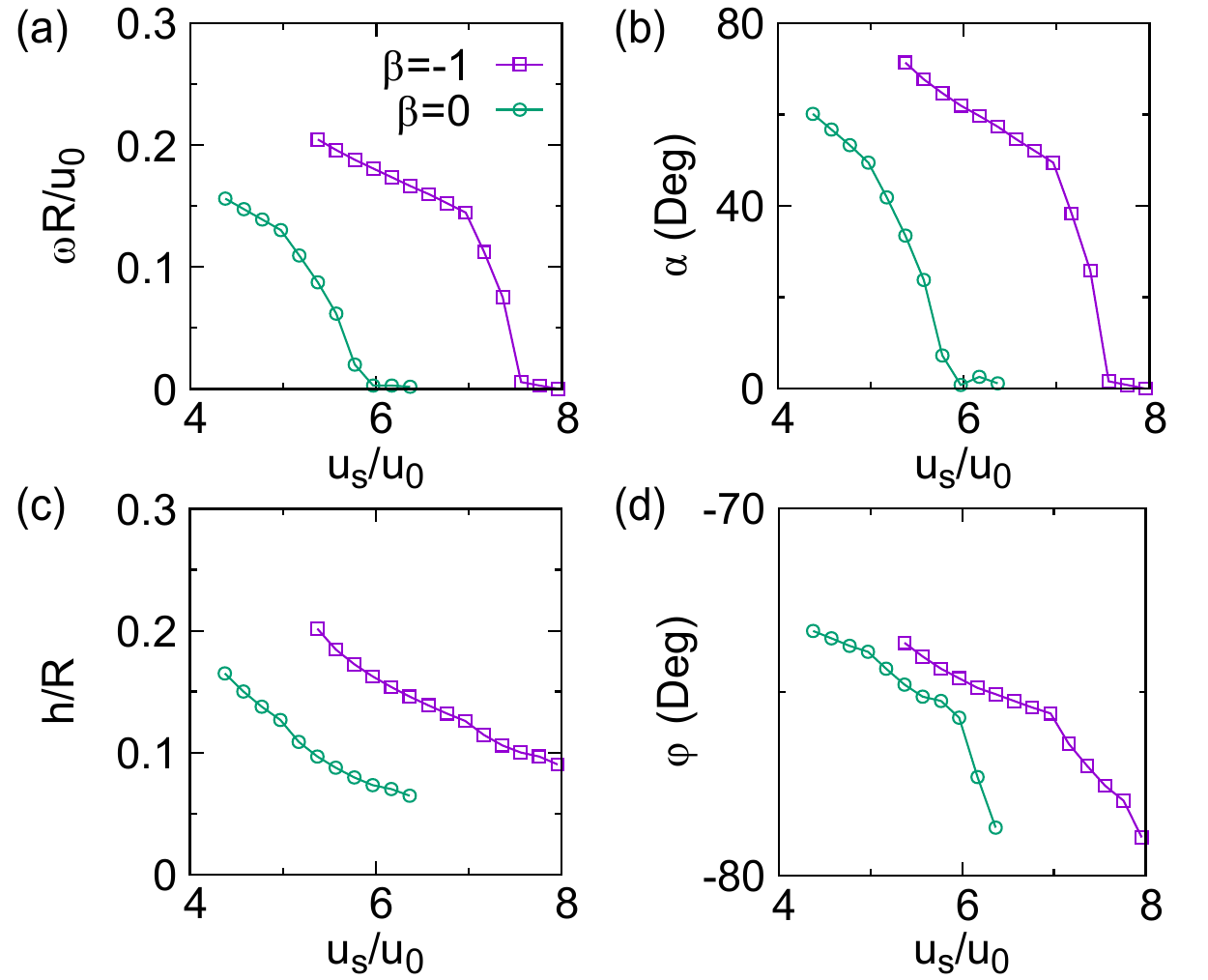}
\caption{\footnotesize {\color{black}(a) The rotation frequency $\omega R/u_0$, (b) the chirality $\alpha$ of the dimer structure, (c) the gap-size $h$ and (d) the orientation angle $\varphi$ for a dimer as a function of the sedimentation strength.}}
\label{dimer}
\end{figure}

\begin{figure}[tbhp!]
\centering
\includegraphics[width=0.5\textwidth]{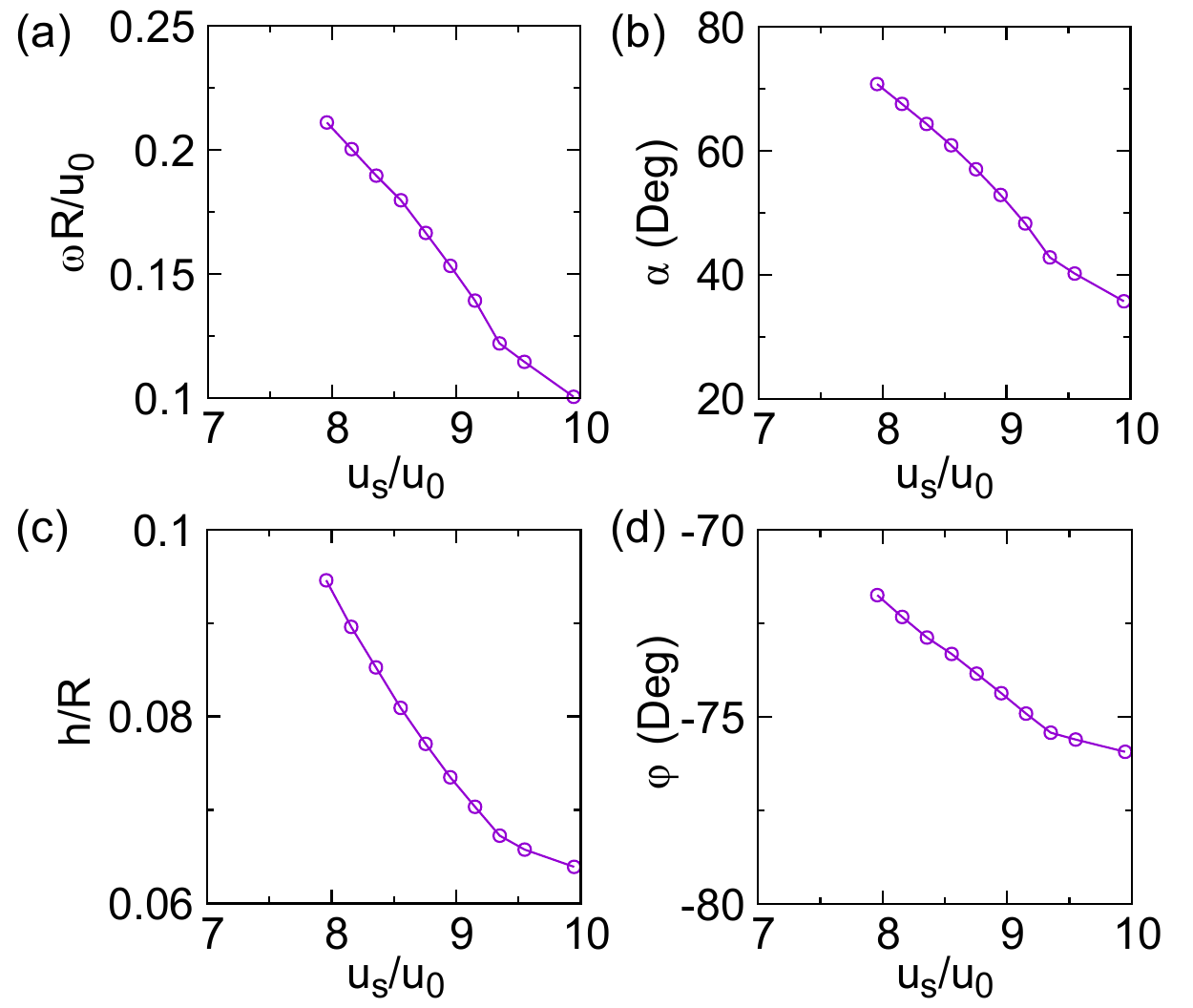}
\caption{\footnotesize {\color{black}(a) The rotation frequency $\omega R/u_0$, (b) the chirality $\alpha$ of the trimer structure, (c) the gap-size $h$ and (d) the orientation angle $\varphi$ for a trimer with $\beta = -1$ as a function of the sedimentation strength.}}
\label{trimer}
\end{figure}

{\color{black} We then consider particle-particle interactions in the floating state for both weak pusher ($\beta=-1$) and pullers ($\beta =+1$) as well as for neutral $\beta = 0$ swimmers.  Interestingly, the stabilisation of hydrodynamically bound particle dimers and trimers are observed (Fig.~\ref{dimer_exp}-\ref{unstable}) for $\beta \leq 0$ swimmers. The stabilisation of dimers is observed for both neutral $\beta = 0$ and $\beta = -1$ swimmers, while the trimers are only observed for $\beta = -1$ pushers.   
The particles are bonded together by purely hydrodynamic interactions. The interplay of the near-field interactions between the particle-particle and the particle-wall makes the particle trap and orient each other, leading to a chiral structure of the particle axes, resulting in a rotational motion of the dimers and trimers in a steady state (Fig.~\ref{dimer_exp} and~\ref{trimer_exp}).}

The stability of the dimer and trimer spinners is related to the wall-particle distance $h$, and can be controlled by the sedimentation strength {\color{black} (Fig.~\ref{dimer} and~\ref{trimer}).}  To study the stability of the dimers and trimers, the particles were initialised in a chiral bound configuration {\color{black} (see e.g. Fig.~\ref{dimer_exp}a and \ref{trimer_exp}a)}, and the sedimentation strength $u_s/u_0$ was varied. {\color{black} No stable hydrodynamically bound structures were observed for a weak sedimentation $u_s/u_0 \lesssim 4.2$. Dimers were observed to be stable when $u_s/u_0 \gtrsim 5.4$ ($h\lesssim 0.2R$) for $\beta=-1$ and $u_s/u_0 \gtrsim 4.2$ ($h\lesssim 0.15R$) for $\beta=0$ (Fig.~\ref{dimer}c), while trimers require $u_s/u_0 \gtrsim 8$ ($h \lesssim 0.1R$) (Fig.~\ref{trimer}c). No stable spinners were found for $\beta = +1$ pullers.

The cluster rotation is due to internal chirality characterised by the angle $\alpha$ between the projection of the vectors from the particle centre to the mass centre of the cluster and the particle orientation, onto the wall plane (see {\it e.g.} Fig~\ref{dimer_exp}a and \ref{trimer_exp}a). Both, the rotation frequency $\omega R/u_0$ and the internal chirality $\alpha$ decrease when the sedimentation is increased (see Fig.~\ref{dimer}a, b and ~\ref{trimer}a, b for  dimers and trimers, respectively). The dimer spinners stop rotating when $h\lesssim 0.07R$ and become achiral. The trimers remain spinning until the computational limit $h \sim 0.05R$ given by the requirement of at least one free fluid node between the particle surface and the wall,  is reached. After this limit the simulations are stopped. In the case of both dimers and trimers, the swimmers tilt slightly away from the wall normal, even after they become achiral (Fig.~\ref{dimer}d and \ref{trimer}d), suggesting that the mutual interactions arising from the self-generated flow fields can compete with the wall induced near-field hydrodynamic interactions.

\begin{figure}[tbhp!]
\centering
\includegraphics[width=0.5\textwidth]{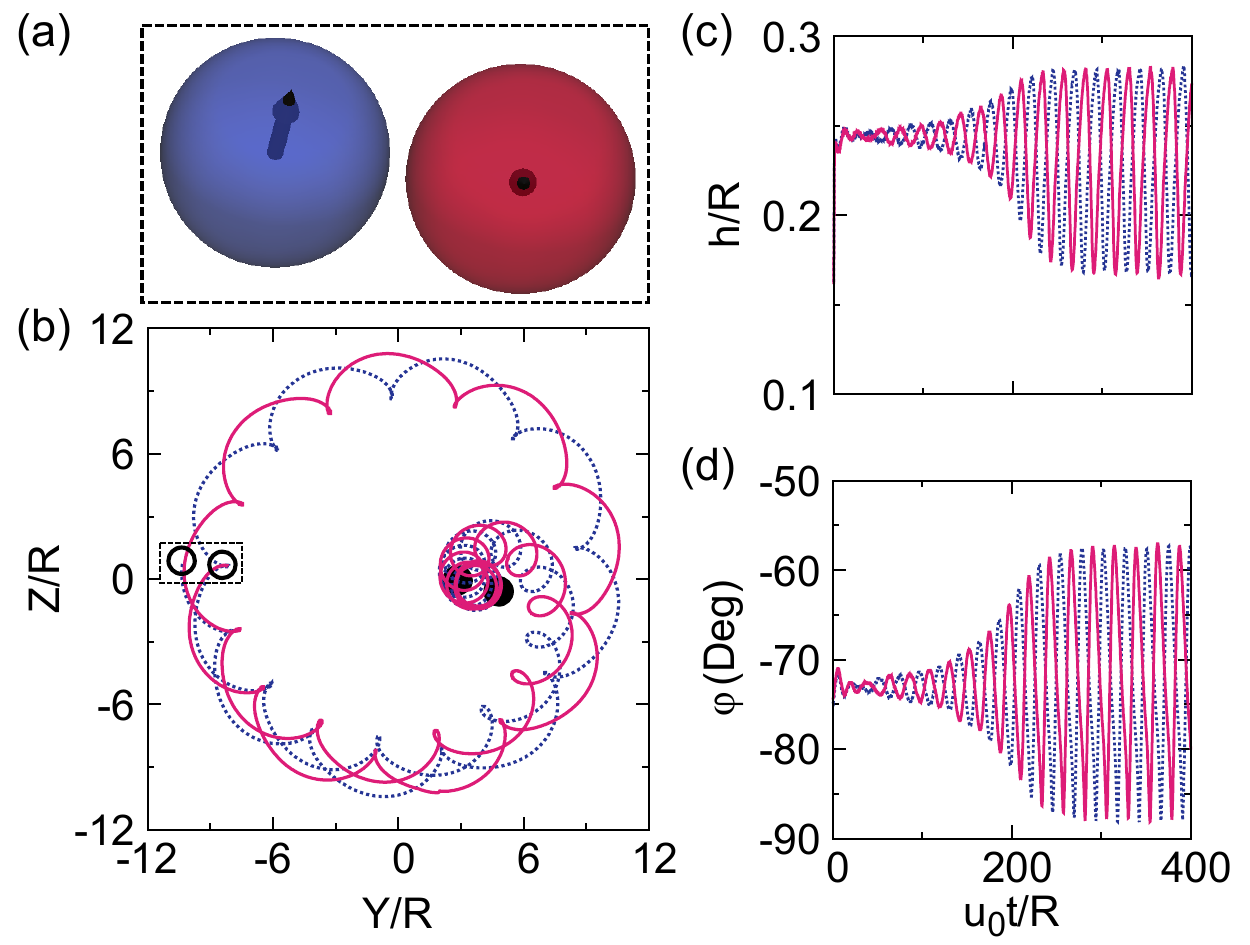}
\caption{\footnotesize {\color{black} A typical case showing a rotor motion of a rotating dimer when the pure spinning motion becomes unstable ($\beta=-1$  and $u_s/u_0=4.8$ ). (a) A snapshot of the configuration of the dimer. The black arrows show the projections of the particle axis in $Y-Z$ plane. (b) The trajectories of the particles in the dimer. The initial positions and final positions are shown by the black solid and hollow circles respectively. The final configuration is given in a. (c) The gap-size $h$ and (d) the orientation angle $\varphi$ of each particle in the dimer are plotted as a function of time $u_0 t/R$.}}
\label{unstable}
\end{figure}

Interestingly, for the dimers near the lower limit of the stability ($4.6 \lesssim u_s/u_0 \lesssim 5.4$), for example ($u_s/u_0 \sim 4.8$), we observe a stark change in the dynamics (Fig.~\ref{unstable}). Now, in addition to the particles spinning around each other, a persistent centre-of-mass movement is observed, where the particle pair settles on a circular trajectory (Fig.~\ref{unstable}b). In this state both the gap size $h$ and the particle orientations $\varphi$ undergo periodic oscillations (Fig.~\ref{unstable}c and d). The oscillations are also internally synchronised. When one of the particles points away from the wall its $h$ increases, while the other simultaneously has a larger inclination angle respect to the wall normal, and its $h$ decreases as schematically shown in Fig.~\ref{unstable}. This leads to a periodic modulation of both $h$ and $\phi$ (Fig.~\ref{unstable}c and d).
}

\subsection{Phase diagram for strong sedimentation}

\begin{figure}[tbhp!]
\centering
\includegraphics[width=0.5\textwidth]{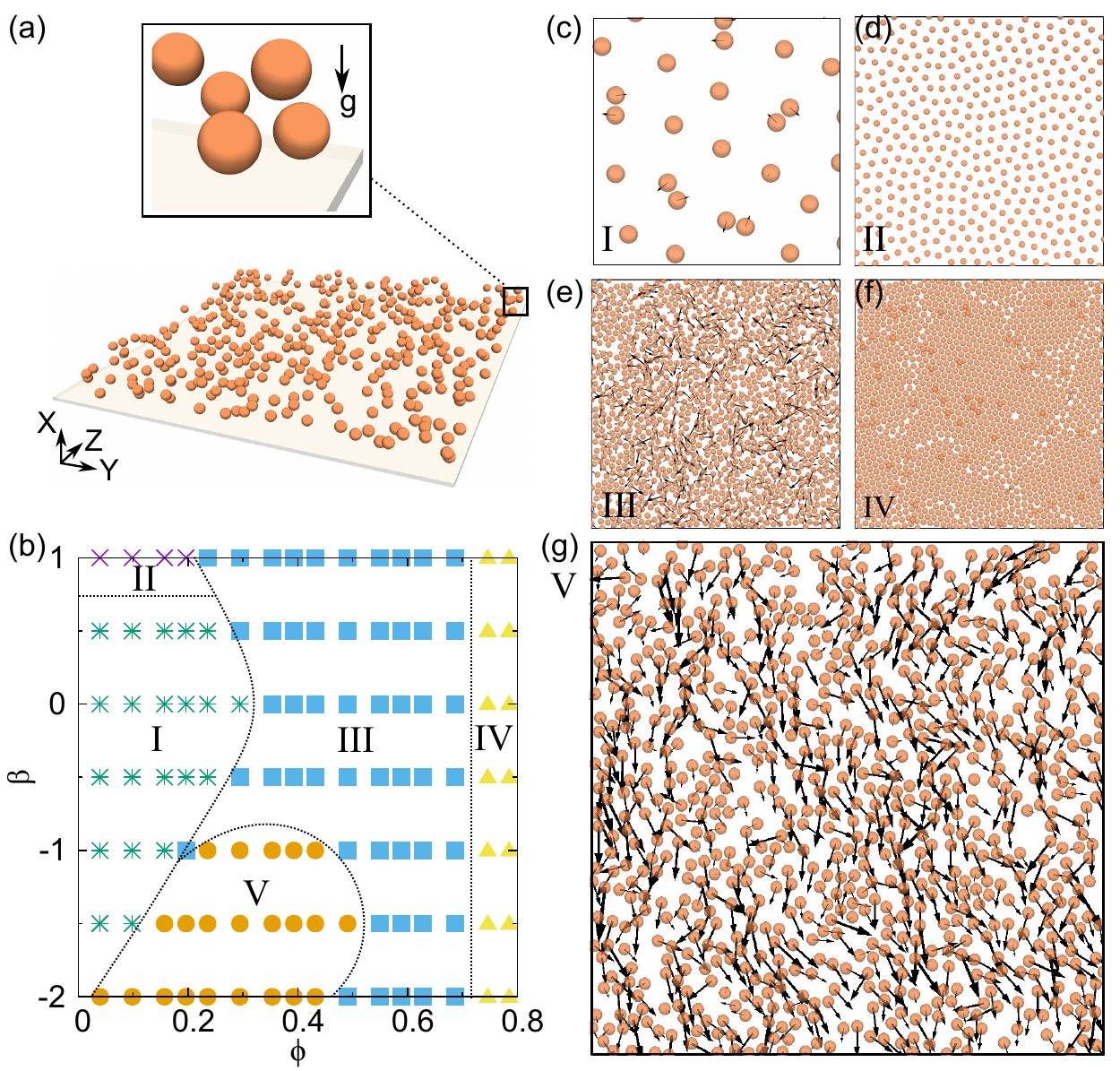}
\caption{\footnotesize (a) Schematic of the simulation set-up. Particles are initially randomly positioned above a solid surface. The sedimentation force $F_s$ is then turned on, leading to the formation of a monolayer of particles near the surface with an area fraction $\phi$ . (b) Different phases observed when the area fraction $\phi$ and the squirmer parameter  $\beta$ are varied. The corresponding configurations for each phase are shown in (c-g). (c) Phase I: rotating dimers. (d) Phase II: isolated particles. (e) Phase III: erratic motion. (f) Phase IV: crystal. (g) Phase V: Polar order. The black arrows in (c-g) show the velocities of the particles. The cases in (c-g) use the simulation box of $L_X = 6R,~L_Y = 80R$ and $L_Z = 80R$.}
\label{phase}
\end{figure}

{\color{black} The above stability and dynamics of hydrodynamically bound small clusters were studied for small to moderate sedimentation strengths ($2 < u_s/u_0 < 10$)  in the absence of external repulsive wall-particle interactions.}
To study the collective dynamics of the squirmers near the confining surface, {\color{black} and to compare with recent predictions~\cite{kuhr2019collective},} we simulate a suspension of $N$ particles with an area fraction of the monolayer $\phi = N \pi R^2/L_YL_Z$ . The squirmers are initialised randomly above the wall (Fig.~\ref{phase}a). A strong sedimentation force is applied on each particle ($u_s/u_0=20$) and an artificial repulsion force is added when the distance between the wall and a particle surface is smaller than $d_c=0.15R$ (Eq.~\ref{repulsion} with $R\approx 8\Delta x$). The balance between these two forces maintains a steady state $h\approx 0.13R$ in all of the simulations.

The Fig.~\ref{phase}b present a state diagram of the observed dynamical states for the area fraction $\phi\in [0,0.8]$ and squirmer parameter $\beta \in [-2, +1]$.
At low area fractions $\phi \lesssim 20$\%, the phase diagram is dominated by a state where hydrodynamically stabilised spinners co-exist with isolated swimmers (state I in Fig.~\ref{phase}b and c; see also Movie S1). Here we observe only the formation of stable dimer spinners, while in the Fig.~\ref{spinner} the formation of stable dimer and trimer spinners was predicted. The difference, most likely, arises from the repulsive  potential interactions between the particles and the wall, setting the the steady state $h~\approx 0.13R$. Fig.~\ref{dimer} and \ref{trimer}, demonstrate that the spinning dynamics is strongly dependent on the steady state $h$ set by the competition between sedimentation and self-propulsion ($u_s/u_0$).
Purely hydrodynamic stabilisation of dimer spinners were observed for $h > 0.1R$, while trimer spinners require $h < 0.1R$ (see Fig.~\ref{dimer}c and \ref{trimer}c).
For $\beta = +1$ pullers, {\color{black} no bound dimers were observed}, and at low $\phi$ a state with isolated particles repelled by the mutual flow fields is observed (state II in Fig.~\ref{phase}b and d).

When the area fraction is increased, the particle-particle interactions destabilise the spinners, and leads to a state where the particles move erratically (state III in Fig.~\ref{phase}b and e). At relatively low volume fractions, the particles form short lived hydrodynamically bound spinners (see example for $\beta = -1$, $\phi=20$\% in Movie S1), while at higher $\phi$ erratic swarms are observed (see $\beta = -1$, $\phi=60$\% in Movie S1).

Reasonably strong pushers $\beta \in [-2,-1]$ are observed to form a polar flowing state, characterised by a large scale collective flow and polar order when $\phi \lesssim 40$\% (state V in Fig.~\ref{phase}b and g). Finally when $\phi > 70$\%, a dense crystal state is observed (state IV in Fig.~\ref{phase}b and f).
The calculated state diagram shows a good qualitative agreement, with a state diagram calculated very recently using MPCD technique~\cite{kuhr2019collective} using similar value for the sedimentation $u_s/u_0 \approx 16.7$ and including thermal effects~\cite{kuhr2019collective}. {\color{black} This suggests that the hydrodynamic interactions are dominant, while thermal fluctuations can affect both the location of the boundaries between the different states and the stability of the observed structures. The main difference between the phase diagram calculated here and in~\cite{kuhr2019collective} is the appearance of the stable spinners at low area fractions (state I in Fig.~\ref{phase}b). The stability of the spinners is strongly dependent of the steady state floating height $h$. It is possible that thermal fluctuations or other interactions, {\it e.g} repulsive (thermodynamic) wall-particle interactions can de-stabilise the spinners.}

\subsection{Polar order state with pushers}
\begin{figure}[tbhp!]
\centering
\includegraphics[width=0.5\textwidth]{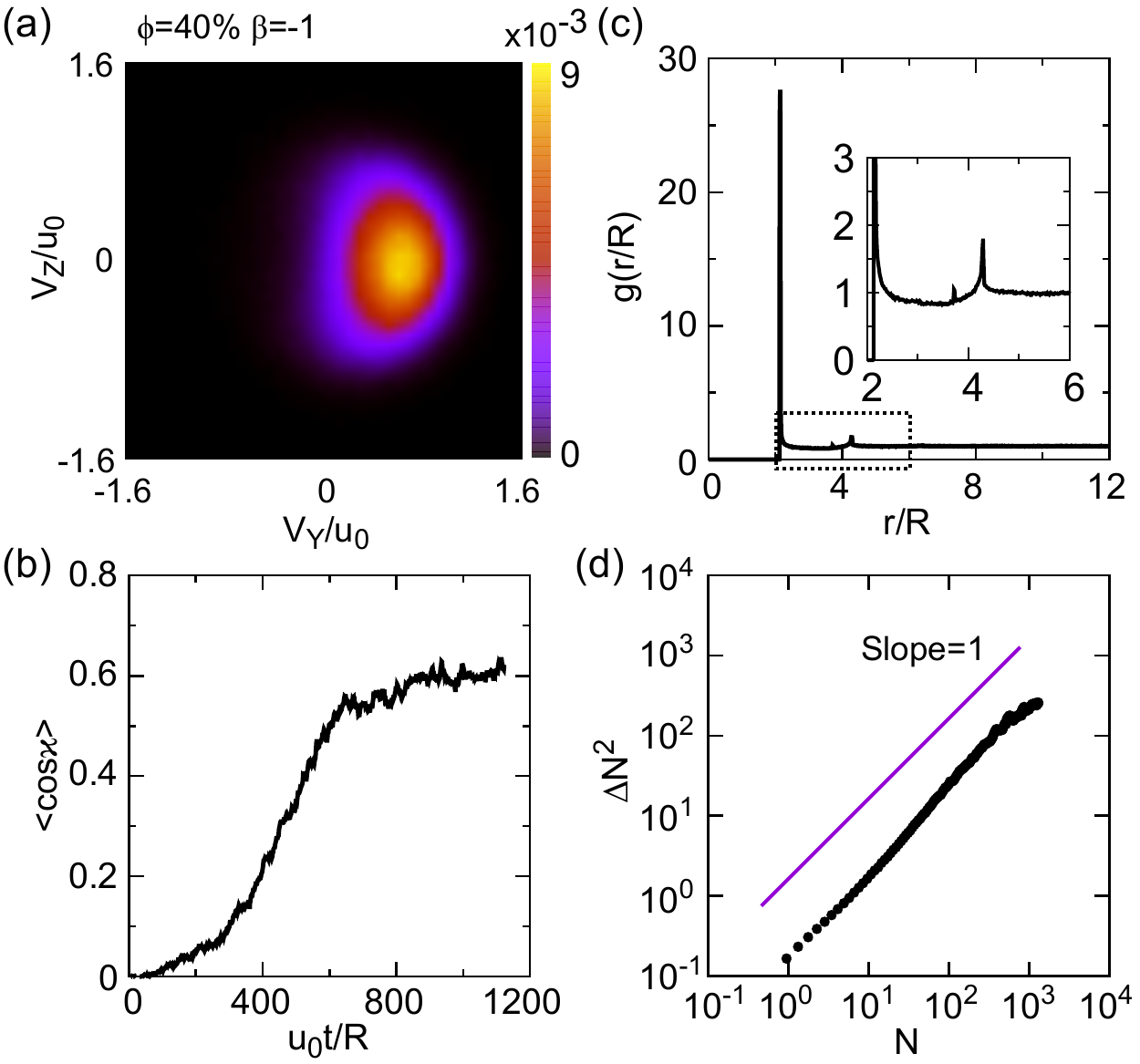}
\caption{\footnotesize Dynamics of particles in a polar state  ($N=3200$, $\phi \approx 40$\% and $L_X = 6R,~L_Y = 160R$ and $L_Z = 160R$.) (a) An example of a two dimensional particle velocity distribution in the polar state. (b) The polar order parameter $\langle \cos\varkappa\rangle$ as a function of time $tu_0/R$. 
The calculated particle (c) pair correlation function $g(r)$ and (d) the number fluctuations $\Delta N^2$ as a function of the average number of particles $N$.}
\label{polar}
\end{figure}

A stable polar order in 3D bulk systems of swimming squirmers have been predicted to occur with pullers~\cite{evans2011orientational,ishikawa2008development,alarcon2013spontaneous}, while typically in pusher suspensions polar order is thought to be unstable. Interestingly, our simulations show stable polar order in the monolayers of reasonably strong {\it pushers} near the confining surface.
To investigate in detail the dynamics of the polar phase, we simulated a larger system $N=3200$ particles with $\phi\approx 40$\% (Fig.~\ref{polar} ).
Starting from random initial orientations, the polar order quickly develops (Fig.~\ref{polar}b). The polar order parameter $\langle \cos\varkappa\rangle$ {\color{black} is calculated from the particle velocities $cos\varkappa_{ij} = \mathbf{v}_{i}\cdot \mathbf{v}_{j}/|v_i v_j|$ and averaged over all the pairs.} Starting from isotropic state it evolves rapidly to reach a steady state value $\langle \cos\varkappa\rangle\approx 0.6$ (Fig.~\ref{polar}b). The steady state velocity probability distribution becomes strongly anisotropic, further signalling a preferential direction of motion, along $Y$-axis in this case (Fig.~\ref{polar}a). The pair-correlation function $g(r)$, shows a strong peak at $r\approx 2R$ and a weaker peak at $r\approx 4R$ (Fig.~\ref{polar}c). 
This can be understood in the terms of the emergence of chaining of particles along a common swimming direction in the polar state (see {\it e.g.} Fig.~\ref{phase}g and Movie S2). Finally, the number fluctuations $\Delta N^2$ show a {\color{black} nearly linear scaling, signalling the absence of giant number fluctuations in the system}, in agreement with the experiments of Quincke rollers in a polar state~\cite{geyer2018sounds}.

\begin{figure}[tbhp!]
\centering
\includegraphics[width=0.5\textwidth]{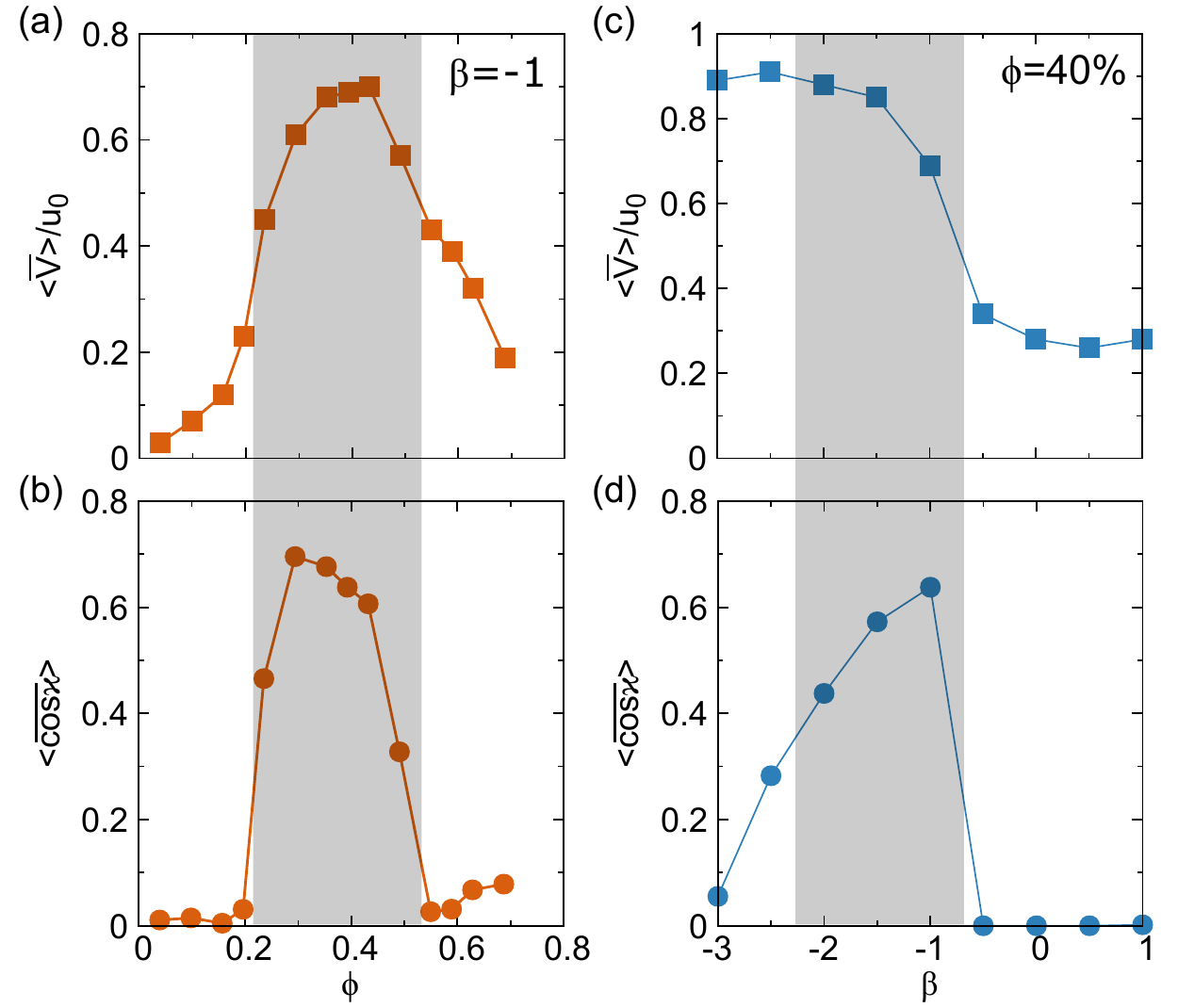}
\caption{\footnotesize (a) The average velocity $\langle \overline{v}\rangle/u_0$  and (b) the polar order parameter $\langle \overline{\cos\varkappa} \rangle$ as a function of the area fraction $\phi$ for $\beta = -1$ pushers. The average velocity and (d) the polar order parameter as a function of $\beta$ for $\phi\approx 40$\%. The shaded areas indicate the range of the polar state.}
\label{polar_phase}
\end{figure}

The origin of the polar order found with pushers in this system is due to the near-field hydrodynamic interactions in the presence of a wall. Near-field hydrodynamic interactions have been predicted to stabilise polar order in monolayers of both weak pushers and pullers interacting through 3-dimensional flow fields~\cite{yoshinaga2017hydrodynamic} .
In our case, the isolated particles are stationary and lean on the surface. The near-field hydrodynamics between the particles promotes small tilts on the particle orientations, leading to a collective movement of the particles. For $\beta=-1$ pushers, a threshold concentration $\phi\approx 20$\% is observed (Fig.~\ref{polar_phase}a and b). Below this value, there is not enough interactions between the swimmers to actuate the collective dynamics and the particles exhibit only single floating particles or rotating dimers (Fig.~\ref{phase}).
We measure {\color{black} the time and space averaged velocity ${\langle\overline{v}\rangle}/u_0$ and the polar order parameter $\langle\overline{\cos\varkappa}\rangle$ in the steady state} for $\beta=-1$ swimmers as a function of the area fraction $\phi$ (Fig.~\ref{polar_phase}a). A stable polar order is observed in the range  $\phi\in [0.2, 0.5]$.
The maximum velocity ${\langle\overline{v}\rangle}_{\mathrm{max}}\approx 0.7u_0$ is observed at $\phi\approx 40$\%, while the maximum polar order $\langle\overline{\cos\varkappa}\rangle \approx 0.7$ takes places at lower area fraction $\phi\approx 30$\%. In order to study the $\beta$ dependence of the polar order, the area fraction was fixed to $\phi=40$\%. In this case, a stable polar order was observed for $-2<\beta -1$ pushers (Fig.~\ref{polar_phase}c and d).
\begin{figure}[tbhp!]
\centering
\includegraphics[width=0.5\textwidth]{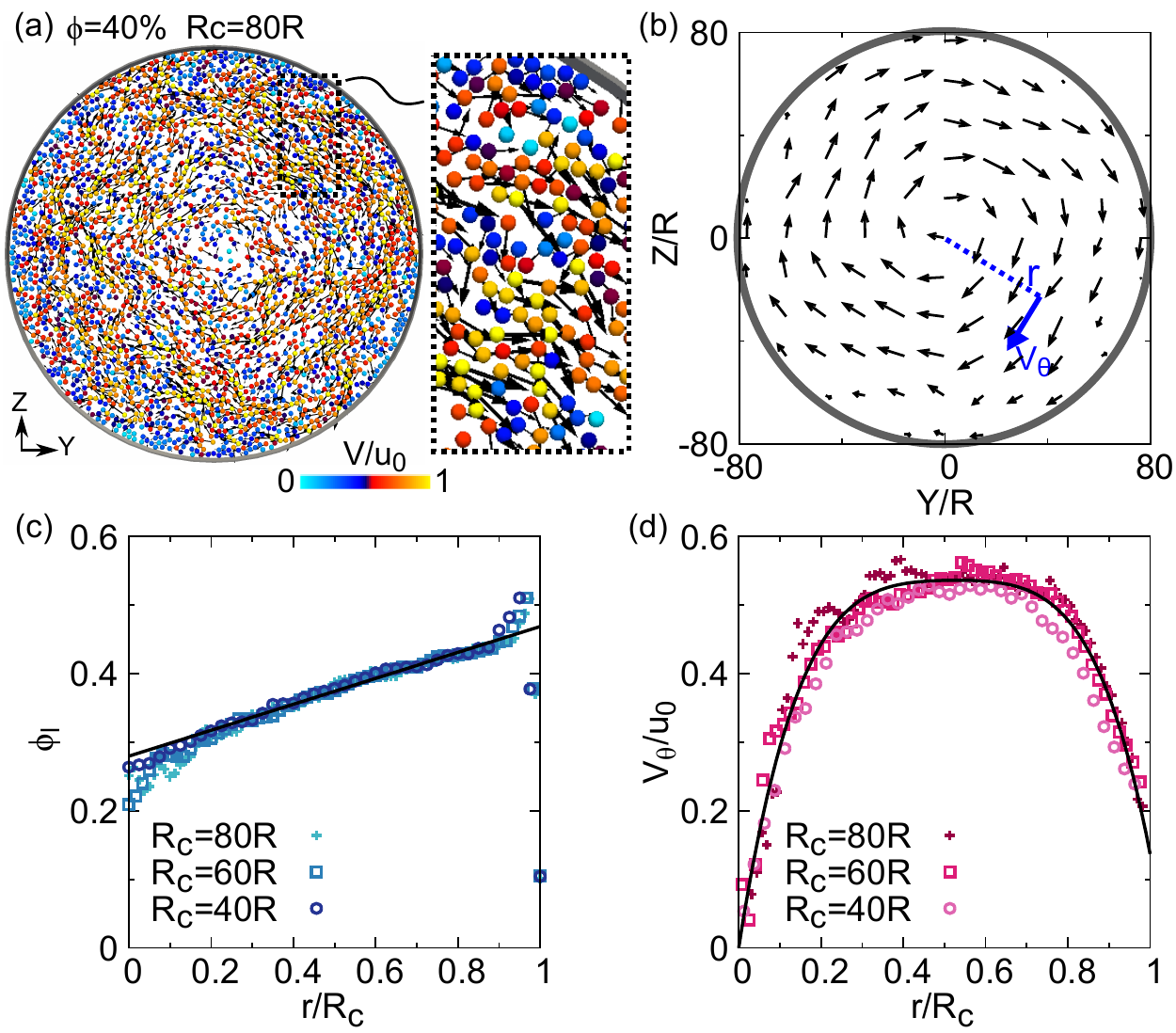}
\caption{\footnotesize (a) A snapshot of particles vortex in cylindrical boundary. The colour code and the arrows show the velocities of the particles. (b) Time-averaged velocity field of the particles. (c) The local area fraction $\phi_l$ along the radius for different size of the cylindrical boundary. (d) The azimuthal velocity $V_\theta/u_0$ along the radius. (The black lines are guide to eyes)}
\label{boundary}
\end{figure}

Finally, we demonstrate that the polar state can survive in the presence of a boundary. When confined in a circular arena on the wall plane {\color{black} with a no-slip boundary condition at the cylindrical boundary}, the particles order themselves into a vortical motion (Fig. ~\ref{boundary}a and b; see also Movie S3). We measure the local density $\phi_l$ along the radial direction (Fig. ~\ref{boundary}c). The local density increases linearly along the radius, and the particles are slightly depleted from the centre of the arena, while slight accumulation is observed at the boundary  {\color{black} similarly to what is observed with colloidal rollers in cylindrical confinement~\cite{bricard2015emergent}. This suggests the observed behaviour is due to collective flow effects, rather than the propulsion mechanism of the individual particles.} 
The time averaged azimuthal velocity $V_\theta/u_0$ shows, that  the collective flow is retained. It is reduced both close to the wall  {$r>0.8R$} and  in the centre of the cylinder {$r<0.2R$}. {\color{black} The velocity must vanish in the middle of the container due to the symmetry of the flow, while near the wall it is reduced due to the interactions with the no-slip wall (Fig.\ref{boundary}d). Between these, a plateau is observed, corresponding to the swimming speeds observed in the swarming state without the circular confinement (Fig.~\ref{polar_phase}a).} 
Interestingly, the observation are not dependent of the cylinder radius; all the simulations using three different radii ($R_{c} = 40,~60,~80R$) collapse on the same curve, as a function of the dimensionless distance $r/R_c$ (Fig.~\ref{boundary}c and d). 

\section{Conclusions}
Using numerical simulations, we have investigated the hydrodynamics of a monolayer of spherical squirmers sedimented on a flat no-slip surface.
At low surface coverage, our results demonstrate the spontaneous formation of hydrodynamically bound spinners, consisting of dimers and trimers. Their stability and rotational frequency can be controlled by the sedimentation strength.
Introducing an external repulsion between the particles and wall, we map out a state diagram at the limit of strong sedimentation speed compared to the swimming speed $u_s/u_0= 20$. We observe the formation of chiral spinners, large scale polar flow for pushers and finally high coverage crystalline state. Finally we test the robustness of the polar order against confinement effects and observe the emergence of particle vortex in a circular confinement.

Our results are generic and crucially require near-field hydrodynamic effects. They should be valid in a wide variety of experimental scenarios such as spherical bacterial swimmers near surfaces~\cite{drescher2009dancing} or artificial swimmers~\cite{moran2017phoretic} provided that hydrodynamic interactions are dominant. This could be realised for example by considering emulsion based swimmers~\cite{thutupalli2018flow} or phoretic Janus swimmers, with a reasonably high squirmer parameter $|\beta| > 1$, such as $\beta \approx -2.45$ recently measured for active Janus swimmers~\cite{campbell2019experimental}.

\begin{acknowledgments}
Z.S. and J.S.L. acknowledge support by IdEx (Initiative d'Excellence) Bordeaux and computational resources from Avakas and Curta clusters.
\end{acknowledgments}

\appendix
\section{Movie descriptions}

Movie S1: Examples of the states observed with $\beta=-1$ pushers (from left to right): hydrodynamically stabilised spinners co-exist with isolated particles ( $\phi=10 \%$, $N = 200$ particles); short lived hydrodynamically bound spinners ($\phi=20 \%$, $N=400$); polar order state ($\phi=40\%$, $N=800$) and erratic swarms  ($\phi=60 \%$, $N=1200$). The computational domain is $L_X=6R,L_Y=80R,L_Z=80R$. The particle radius is $R=8 \Delta X$. \\
\\
Movie S2: Polar state with $\beta=-1$ pushers ($N=3200$; $\phi\approx 40$\%). The computational domain is $L_X=6R,L_Y=160R,L_Z=160R$. The particle radius is $R=8 \Delta X$.\\
\\
Movie S3: The formation of a particle vortex in a cylindrical confinement with a radius $R_c\approx 60R$ ($\beta = -1$, $N=1440$, $\phi\approx 40$\%).

\bibliography{ref}





\end{document}